\documentclass{PoS}

\title{On qualitative aspects of the choice of factorization schemes at NLO}

\ShortTitle{On qualitative aspects of the choice of factorization schemes at NLO}

\author{\speaker{Karel KOLAR}\\
        Institute of Physics, Prague\\
        E-mail: \email{kolark@fzu.cz}}

\abstract{Although the choice of a factorization scheme is as important as the choice
of a factorization scale, the dependence of theoretical predictions (at finite order)
on the choice of a factorization scheme has been little investigated. This is due to
the fact that the freedom in the choice of a factorization scheme is enormous, even
at NLO. One of the reason why to study factorization schemes is the possible
exploitation of the freedom in their choice in the construction of NLO Monte Carlo
event generators with NLO initial state parton showers. However, the ZERO factorization
scheme, which should be optimal for such Monte Carlo event generators, has turned out
to be practically inapplicable although it appears at first sight as reasonable.
A detailed analysis has then shown that if some given NLO splitting functions do
not satisfy a certain nontrivial condition, then the corresponding factorization
scheme has some restrictions on its practical applicability. Relevant technical
details of the discussed issues are the content of this short contribution.}

\FullConference{European Physical Society Europhysics Conference on High Energy
Physics\\
                 July 16-22, 2009\\
                 Krakow, Poland}

\begin{document}

\section{The freedom in the choice of a factorization scheme at NLO}

To illustrate the issue of the freedom in the choice of a factorization scheme, consider
a structure function $F\!\!\left(x, Q^2 \right)$ which is expressed by the formula
\begin{equation}
  F\!\!\left( x, Q^2 \right) = \sum_i \int_x^1 {{\rm d}y\over y}\,\, C_i \!\left( {x\over y},
  Q^2, M, {\rm FS}\right) D_i \left(y, M, {\rm FS}\right), \label{eq1}
\end{equation}
where $C_i \!\left( x, Q^2, M, {\rm FS}\right)$ stands for the corresponding coefficient
functions and $D_i \left( x, M, {\rm FS} \right)$ represents the parton distribution
functions. The coefficient functions are fully calculable within the framework of
perturbative QCD and can thus be expanded in powers of the QCD coupling parameter
$a \equiv \alpha_{\rm s} / \pi$
\begin{equation}
  C_i \!\left( x, Q^2, M, {\rm FS}\right) = C^{(0)}_i \delta (1-x) + a(\mu)\, C^{(1)}_i
  \!\!\left( x, Q^2, M, {\rm FS}\right) + {\cal O}\!\left( a^2(\mu) \right) . \label{eq2}
\end{equation}
The parton distribution functions satisfy the evolution equations
\begin{equation}
  {{\rm d} D_i \left(x, M, {\rm FS}\right) \over {\rm d} \ln M} = a(M) \sum_j \int_x^1
  {{\rm d}y\over y}\,\, P_{ij}\!\left( {x\over y}, M, {\rm FS}\!\right) D_j \left(
  y, M, {\rm FS}\right),
\end{equation}
where the splitting functions $P_{ij}\!\left( x, M, {\rm FS}\right)$ can be expanded
in powers of $a(M)$
\begin{equation}
  P_{ij}\left(x, M, {\rm FS}\right) = P^{(0)}_{ij}(x) + a(M)\,P^{(1)}_{ij}\!\left(x, {\rm FS}
  \right) + {\cal O}\!\left( a^2(M) \right) . \label{eq4}
\end{equation}
The higher order splitting functions $P^{(k)}_{ij}\!\left(x, {\rm FS}\right)\!,\,\,\, k \geq 1$,
which can be chosen at will, can be used for labeling factorization schemes. At NLO,
a factorization scheme is fully specified by the corresponding NLO splitting functions
$P^{(1)}_{ij}\!(x)$. If the series (\ref{eq2}) and (\ref{eq4}) are summed to all orders, then
the structure function $F\!\!\left(x, Q^2 \right)$ is independent of the renormalization scale
$\mu$, of the factorization scale $M$ and of the factorization scheme FS, however, {\em finite\/}
order theoretical predictions for the structure function {\em depend\/} on these unphysical quantities.

Changing the factorization scheme of the parton distribution functions is described by the formula
\begin{equation}
  D_i \left( x, M, {\rm FS} \right) = D_i \left( x, M, {\rm FS_0} \right) +
  a(M)\sum_j \int_x^1 {{\rm d}y\over y}\,\, T^{(1)}_{ij}\!\! \left( {x\over y}, {\rm FS},
  {\rm FS_0}\!\right) D_j \left( y, M, {\rm FS_0}\right) + {\cal O}\!\left( a^2(M) \right) .
\end{equation}
The Mellin moments of the matrix function $T^{(1)}_{ij}\! \left(x, {\rm FS_1}, {\rm FS_2}\right)$
are determined by the following matrix equation
\begin{equation}
  \left[ \mathbf{T}^{(1)}(n, {\rm FS_1}, {\rm FS_2}), \,\mathbf{P}^{(0)}(n) \right] - b\,
  \mathbf{T}^{(1)}(n, {\rm FS_1}, {\rm FS_2}) = \mathbf{P}^{(1)}(n, {\rm FS_1})
  - \mathbf{P}^{(1)}(n, {\rm FS_2}) . \label{eq6}
\end{equation}
The solution of the preceding equation can be expressed in analytic form, however, the Mellin
inversion to $x$--space has to be calculated numerically in general. The transformation formula
for the NLO coefficient functions reads
\begin{equation}
  C^{(1)}_i \!\!\left( x, Q^2, M, {\rm FS}\right) = C^{(1)}_i \!\!\left( x, Q^2, M_0, {\rm FS_0}
  \right) + \sum_j C^{(0)}_j \!\left( P^{(0)}_{ji}(x) \ln {M_0 \over M} + T^{(1)}_{ji}\!
  \left( x, {\rm FS_0}, {\rm FS}\right) \right) .
\end{equation}

\section{The ZERO factorization scheme}

The so--called ZERO factorization scheme is defined by the condition that the NLO splitting
functions $P^{(1)}_{ij}\!(x)$ vanish. The NLO initial state parton showers are thus formally
identical to the LO ones in this scheme and therefore can be generated and attached to NLO QCD
cross-sections by the standard algorithms. Using the ZERO factorization scheme, we can thus
obtain NLO Monte Carlo event generators that involve NLO initial state parton showers without
any change in the current algorithms.

\begin{figure}
\begin{center}
  \includegraphics[angle=90,width=0.76\textwidth]{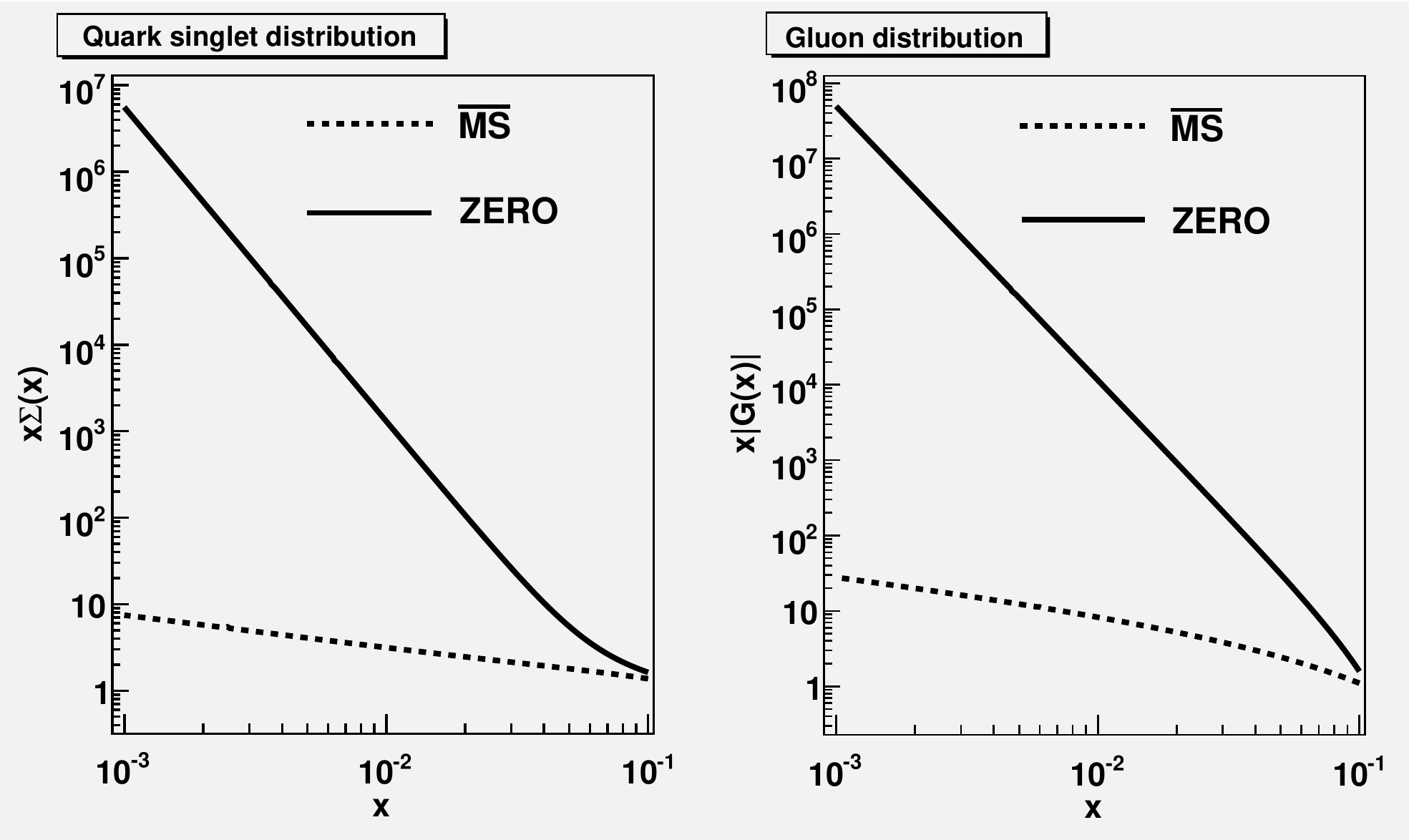}
  \caption{Comparison of the ZERO and $\overline{\rm MS}$ singlet parton distributions at $M = 50
  \, \, {\rm GeV}$. The $\overline{\rm MS}$ distributions were obtained by evolving the starting
  distributions of the MRST98 set \cite{mrst} with the fixed number of active flavours $n_{\rm f} = 3$
  (only light flavours are taken into account). The ZERO distributions were calculated numerically
  from the $\overline{\rm MS}$ ones. Note that the gluon distributions are plotted in their absolute
  value because the ZERO gluon distribution is negative in the displayed region. The point where
  the ZERO gluon distribution changes the sign is close to $x = 0.1$.}
  \label{fig1}
\end{center}
\end{figure}

To obtain the coefficient functions and the parton distributions in the ZERO factorization scheme,
we have to calculate the matrix function $T^{(1)}_{ij}\!\!\left(x, \overline{\rm MS}, {\rm ZERO}
\right)$ \footnote{According to the equation (\ref{eq6}), the matrix function $T^{(1)}_{ij}
\!\!\left(x, {\rm ZERO}, \overline{\rm MS}\right)$, which is required for the transformation of
the parton distributions, is equal to $-T^{(1)}_{ij}\!\!\left(x, \overline{\rm MS}, {\rm ZERO}
\right)$.}. This was performed for three and four effectively massless flavours. The obtained
results are surprising because for $x \lesssim 0.1$
\begin{equation}
  T^{(1)}_{ij}\!\!\left( x, \overline{\rm MS}, {\rm ZERO} \right)\, \approx\, C_{ij} x^{-\xi}
  \quad\; {\rm with} \;\;\; \xi (n_{\rm f} = 3) \doteq 4.63 \;\;\; {\rm and} \;\;\;
  \xi (n_{\rm f} = 4) \doteq 3.85 \, .
\end{equation}
The ZERO coefficient functions and the ZERO parton distributions diverge for low $x$ in a similar
way as the functions $T^{(1)}_{ij}\!\!\left(x, \overline{\rm MS}, {\rm ZERO} \right)$. The ZERO
singlet parton distributions are plotted for $x \in (10^{-3}, 10^{-1})$ in Figure \ref{fig1}.
The divergent terms cause problems in numerical calculations, and moreover, it is likely that
the mutual cancellation of the divergent terms in expressions for physical quantities, such as
(\ref{eq1}), is incomplete at NLO, which can lead to {\em unreasonable\/} theoretical
predictions. From the practical point of view, the ZERO factorization scheme is thus {\em in
general\/} inapplicable, but it can be proven that there are no problems with its applicability
in the non--singlet case.

\section{The condition of applicability}

A given factorization scheme obviously has no restrictions on its practical applicability
if the corresponding coefficient functions, cross-sections at parton level and parton
distribution functions behave for low $x$ in a similar way as the $\overline{\rm MS}$
ones. If we consider only such factorization schemes for which the corresponding matrix
$\mathbf{P}^{(1)}(x)$ has the same structure as the $\overline{\rm MS}$ one, that is
\begin{equation}
\begin{array}{rccclrcccl}
  P^{(1)}_{q_i q_j}(x) & = & P^{(1)}_{\bar q_i \bar q_j}(x) & = &
  \delta_{ij} P^{(1)V}_{qq}\!(x) + P^{(1)S}_{qq}(x), \qquad &
  P^{(1)}_{q_i G}(x) & = & P^{(1)}_{\bar q_i G}(x) & = & P^{(1)}_{qG}(x),
  \\[0.3cm] P^{(1)}_{q_i \bar q_j}(x) & = & P^{(1)}_{\bar q_i q_j}(x) & = &
  \delta_{ij} P^{(1)V}_{q\bar q}\!(x) + P^{(1)S}_{qq}(x), \qquad &
  P^{(1)}_{G q_i}(x) & = & P^{(1)}_{G \bar q_i}(x) & = & P^{(1)}_{Gq}(x),
\end{array}
\end{equation}
then a detailed analysis of the solution of the equation (\ref{eq6}) shows that
the preceding condition of applicability is fulfilled if and only if the Mellin
moments of the corresponding NLO splitting functions $P^{(1)V}_{qq}\!(n)$,
$P^{(1)V}_{q\bar q}\!(n)$, $P^{(1)S}_{qq}(n)$, $P^{(1)}_{qG}(n)$,
$P^{(1)}_{Gq}(n)$ and $P^{(1)}_{GG}(n)$ satisfy the equation
\begin{eqnarray}
  & \!\!\!\!\!\! \!\!\!\!\!\! & \,\,\,\,\,\,\,\, P^{(0)}_{Gq}(n) \left( P^{(0)}_{qq}(n) -
  P^{(0)}_{GG}(n) - b \right) \left( P^{(1)}_{qG}(n) - P^{(1)}_{qG}(n, \overline{\rm MS})
  \right) + \nonumber\\ & \!\!\!\!\!\! \!\!\!\!\!\! & + \, P^{(0)}_{qG}(n) \left(
  P^{(0)}_{qq}(n) - P^{(0)}_{GG}(n) + b \right) \left( P^{(1)}_{Gq}(n) - P^{(1)}_{Gq}(n,
  \overline{\rm MS}) \right) - \nonumber\\ & \!\!\!\!\!\! \!\!\!\!\!\! & - \, 2
  P^{(0)}_{qG}(n) P^{(0)}_{Gq}(n) \left( P^{(1)V}_{qq}\!(n) + P^{(1)V}_{q\bar q}\!(n) +
  2n_{\rm f} P^{(1)S}_{qq}(n) - P^{(1)}_{GG}(n)\, - \right. \nonumber\\
  & \!\!\!\!\!\! \!\!\!\!\!\! & \left. - \, P^{(1)V}_{qq}\!(n, \overline{\rm MS}) -
  P^{(1)V}_{q\bar q}\!(n, \overline{\rm MS}) - 2n_{\rm f} P^{(1)S}_{qq}(n,
  \overline{\rm MS}) + P^{(1)}_{GG}(n, \overline{\rm MS}) \right) = 0
\end{eqnarray}
for $n \in {\cal N}$ where
\begin{equation}
  {\cal N}_{n_{\rm f} = 3}\, = \,\left\{ 1.7329, 4.6306 \right\} ,\quad
  {\cal N}_{n_{\rm f} = 4}\, = \,\left\{ 1.7995, 3.8458 \right\} ,\quad
  {\cal N}_{n_{\rm f} = 5}\, = \,\left\{ 1.9001, 3.1798 \right\} .
\end{equation}

\section{Summary and conclusion}

It has been shown that not all NLO splitting functions that appear at first sight
as reasonable specify practically applicable factorization schemes. The practical
applicability of a factorization scheme is assured if the corresponding NLO splitting
functions satisfy some nontrivial condition, which can be easily formulated in
the space of Mellin moments. This condition is unfortunately not satisfied in
the ZERO factorization scheme, which would otherwise be optimal for NLO Monte
Carlo event generators. Hence, searching for a suitable factorization scheme
which is close to the ZERO factorization scheme and satisfies the condition
of applicability has already been started.

\acknowledgments{\small The author would like to thank J. Ch\'yla for careful reading
of this contribution and valuable suggestions. This work was supported by the projects
LC527 of Ministry of Education and AVOZ10100502 of the Academy of Sciences of
the Czech Republic.}

\end{document}